\documentclass[sigconf]{acmart}
\AtBeginDocument{%
  }

\emergencystretch=\maxdimen
\usepackage{algorithm}
\usepackage{algorithmic}
\usepackage{amsmath}

\begin{document}

\title{Similarity-Based Assessment of Computational Reproducibility in Jupyter Notebooks}


\author{A S M Shahadat Hossain}
\email{ahr8v@missouri.edu}
\orcid{0009-0008-3229-8821}
\affiliation{%
  \institution{University of Missouri}
  \city{Columbia}
  \state{Missouri}
  \country{USA}
}

\author{Colin Brown}
\email{colinjbrown@niu.edu}
\orcid{0000-0002-2214-2726}
\affiliation{%
  \institution{Northern Illinois University}
  \city{DeKalb}
  \state{Illinois}
  \country{USA}
}

\author{David Koop}
\email{dakoop@niu.edu}
\orcid{0000-0002-4422-6162}
\affiliation{%
  \institution{Northern Illinois University}
  \city{DeKalb}
  \state{Illinois}
  \country{USA}
}

\author{Tanu Malik}
\email{tanu@missouri.edu}
\orcid{0009-0007-9656-727X}
\affiliation{%
  \institution{University of Missouri}
  \city{Columbia}
  \state{Missouri}
  \country{USA}
}

\renewcommand{\shortauthors}{Hossain et al.}

\begin{abstract}
Computational reproducibility refers to obtaining consistent results when rerunning an experiment. Jupyter Notebook, a web-based computational notebook application, facilitates running, publishing, and sharing computational experiments along with their results. However, rerunning a Jupyter Notebook may not always generate identical results due to various factors, such as randomness, changes in library versions, or variations in the computational environment. This paper introduces the Similarity-based Reproducibility Index (SRI) – a metric for assessing the reproducibility of results in Jupyter Notebooks. SRI employs novel methods developed based on similarity metrics specific to different types of Python objects to compare rerun outputs against original outputs. For every cell generating an output in a rerun notebook, SRI reports a quantitative score in the range [0, 1] as well as some qualitative insights to assess reproducibility. The paper also includes a case study in which the proposed metric is applied to a set of Jupyter Notebooks, demonstrating how various similarity metrics can be leveraged to quantify computational reproducibility. 
\end{abstract}




\maketitle

\section{Introduction}

In recent years, computational reproducibility has become one of the major concerns among various research communities. It has gained increasing attention since the National Academies of Sciences, Engineering, and Medicine published the report titled ``Reproducibility and Replicability in Science'' \cite{nas_report} in 2019. Because better reproducibility can help save energy, financial cost, and time to reproduce an outcome, it is not only computer science but also other domains such as geoscience \cite{Steventon2022-yz}, management science \cite{Fisar2024-uo}, and social sciences \cite{Moody2022-di} that are aware of the need to form a reproducibility culture. 

Ideally, rerunning a scientific experiment is expected to generate outcomes identical to the results observed in the original experiment. However, practically, this expectation does not always hold due to various factors such as environmental catalysts, changes in methods of recording the results, or even changes in the measuring units. A computational experiment is also not much different from this manner. Rerunning the same code in the same computational environment, even with the same data, may generate results that are different from the original ones. These changes in results come in various forms. Sometimes, these changes are subtle and are even difficult to perceive, however, sometimes they can make a significant difference in the result interpretation. To understand how such changes impact the reproducibility of results, it is crucial to have a standard metric to measure the reproducibility that goes beyond the simple binary judgment of two results being completely ``identical'' or ``different''.

Jupyter Notebook is a popular web-based application that offers interactive environments for computational experiments. It allows users to divide the codes into smaller execution blocks called cells to run them separately and save the outputs. Jupyter Notebooks allow users to run and rerun the cells in any order. Though these types of flexibility facilitates the rerunnability of codes, they do not guarantee that the results will be reproduced. The results may rather change in rerun notebooks for 
various reasons, such as randomness, different computing environments, or changed library versions. There are a few comprehensive works \cite{Pimentel2019-wh, Pimentel2021-ai, Wang2021-gu, Samuel2023-sm, ahmad2022reproducible} on the reproducibility of Jupyter Notebooks. However, a standard metric to measure the reproducibility is still absent.

In this paper, we propose a new metric called Similarity-based Reproducibility Index (SRI) to assess the reproducibility of results in Jupyter Notebooks. While the existing practices of evaluating reproducibility solely depend on a deterministic match, SRI is a combination of both quantitative and qualitative evaluation of reproducibility. It is based on the assumption that a higher similarity between two results makes them more reproducible, even if they are not identical. The quantitative evaluation of SRI leverages similarity metrics to compare two notebook cell outputs to generate a reproducibility score in the range [0, 1]. On the other hand, the qualitative evaluation makes the metric more comprehensive by providing useful insight regarding various data definition aspects of the outputs.

We first applied SRI on the cell outputs extracted from notebooks and their rerun versions. After that, we accumulated all the cell-wise SRI results of the entire notebook into a notebook-wise SRI. We dumped the notebook-wise SRI results as JSON (JavaScript Object Notation) structures for better operability. Finally, we performed a case study on 15 notebooks where we generated the JSON structures for each notebook and visualized them using a web browser to cross-check the evaluation results with the notebook cell outputs and analyze the performance of the proposed metric.

The rest of the paper is organized as follows: section \ref{definitions} defines necessary terminologies with examples, section \ref{methodology} presents the methodology of the proposed metric, and section \ref{experiments} applies the metric on Jupyter Notebooks to analyze its performance. Finally, section \ref{literature} discusses related work, followed by the conclusion in section \ref{conclusion}.

\section{Definitions}
\label{definitions}
\subsection{Cells in Jupyter Notebooks}

A cell is the primary building block and modular unit of a notebook. It facilitates modularity by allowing users to break down a computational process into discrete, manageable steps that can be executed independently or in sequence. There are three types of cells in Jupyter Notebooks:
\begin{itemize}
    \item \textbf{Code cells}: Being the most prominent type of cell in notebooks, a code cell is used to write code. Code cells automatically support smart code formatting, such as syntax highlighting and indentations.  When this type of cell is executed, the code written in it is sent to the kernel responsible for running the code. If the code generates any results, then the result is displayed as the output of the respective code cell.
    \item \textbf{Markdown cells}: This type of cell facilitates documentation of the code. The notebook author can use markdown cells to describe the code or the computational processes using natural texts, rich text, and even markup languages. The cells support text formatting using HTML and standard LaTeX notations.
    \item \textbf{Raw cells}: This type of cell is the least commonly used one in notebooks. Raw cells are used to write raw codes for the notebook to be rendered or converted into different formats, such as HTML or LaTeX. Raw cells are not evaluated by notebooks.
\end{itemize}

\begin{figure}
  \centering
  \includegraphics[width=\linewidth]{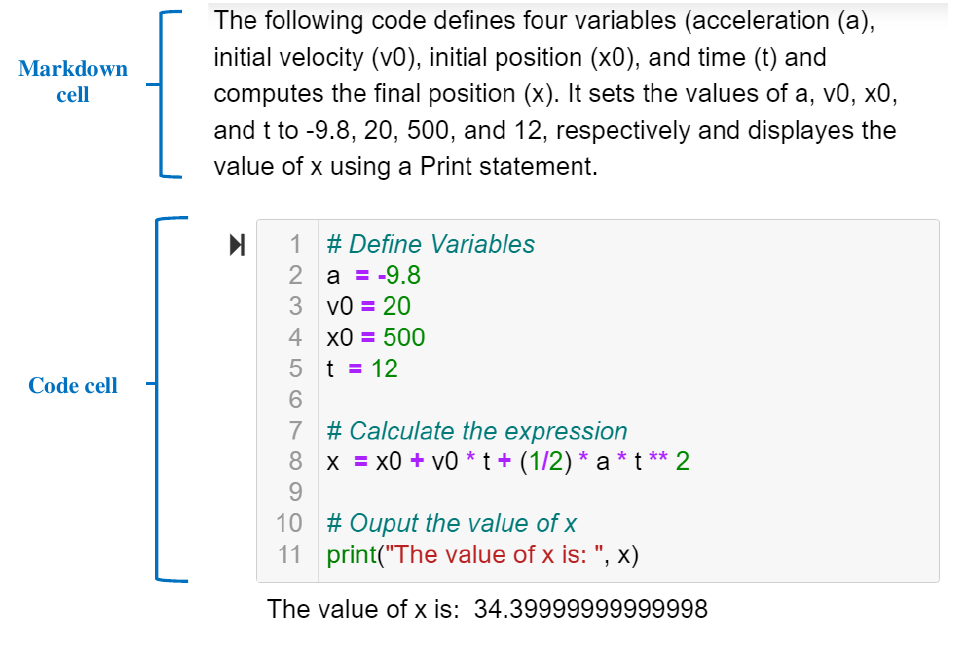}
  \caption{Code Cell and Markdown Cell in a Jupyter Notebook.}
  \label{cell_types}
\end{figure}

Figure \ref{cell_types} presents an example of a code cell and a markdown cell.

\subsection{Types of Code Cell Outputs}
\label{types_of_outputs}

Jupyter Notebook supports various types of outputs. Among them, the major types of outputs observed as the code cell outputs are as follows:

\begin{figure}
  \centering
  \includegraphics[width=\linewidth]{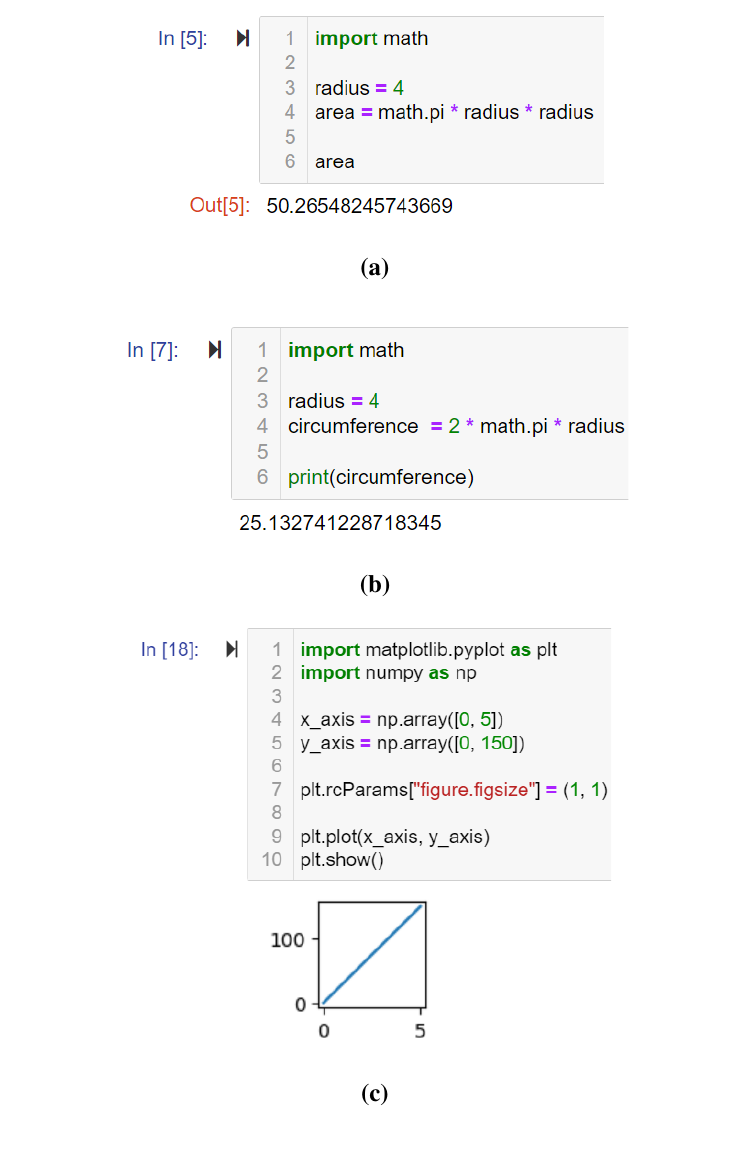}
  \caption{Different Types of Code Cell Outputs in a Jupyter Notebook (a) \textit{execute\_result} (b) \textit{stream} (c) \textit{display\_data}}
  \label{output_types}
\end{figure}

\begin{itemize}
    \item \textbf{stream outputs}: This type of outputs are outputs typically displayed using the \textit{print} statement. The most common MIME types under this category of outputs are \textit{text/plain} and \textit{text/html}.

    \item \textbf{display\_data outputs}: In the JSON format of a notebook, \textit{display\_data} represents the rich type display outputs such as images.  Among different formats of images, \textit{image/png} is the most frequent one observed. Besides, \textit{image/jpeg} and \textit{image/svg} are some of the other MIME types of \textit{display\_data} outputs.

    \item \textbf{execute\_result outputs}: Jupyter Notebook offers a useful feature of writing the name of an object at the end of a code cell to display it without the need to use any \textit{print} statement. The outputs displayed in such cases are generally stored as \textit{execute\_result} outputs in JSON format. 

    \item \textbf{error outputs}: This type of output represents errors due to the failed execution of the codes in a code cell.
    
\end{itemize}

Figure \ref{output_types} illustrates different types of code cell outputs in a Jupyter Notebook.

\section{Methodology}
\label{methodology}

The Similarity-based Reproducibility Index (SRI) is the proposed metric to approximate reproducibility in Jupyter Notebooks. As the name suggests, this metric prioritizes similarity-based comparison between two results. An SRI for a notebook code cell generating outputs is a collection of a quantitative reproducibility score in the range $[0, 1]$ and several qualitative insights regarding the similarity aspects. On the other hand, an SRI for the entire notebook is a JSON (JavaScript Object Notation) structure that contains the names of the original notebook and the rerun notebook and a list of cell-wise SRI results followed by an average of the all cell-wise reproducibility scores. It keeps track of the cell execution IDs for the cells in both original and rerun notebooks. It also includes all the results if one output pair is compared using multiple methods. For example, a \textit{stream} output, which is a plain text, can sometimes be compared as list outputs besides string comparisons. In that case, a cell-wise SRI keeps records of both the actual object type and the converted object type. 

To apply SRI, first, we parse different types of code cell outputs (as discussed in section \ref{types_of_outputs}) from both the original notebooks and their rerun versions. For the \textit{execute\_result} outputs, we determined and added their specific data type to the cell metadata while rerunning the notebooks. After that, we apply different methods to the outputs based on this data type information. As the \textit{stream} outputs are simply rendered as plain texts, we determine their string similarity to evaluate the reproducibility. Similarly, we apply image similarity metrics on the \textit{display\_data} outputs rendered as images. We create A JSON-compatible Python dictionary structure to accumulate all the cell-wise SRI results along with the respective cell execution IDs of the original cell and rerun cell. We also calculate an average of all the cell-wise reproducibility scores to use as the overall reproducibility of the notebook. Finally, we convert the SRI result for an entire notebook into a JSON structure for better interoperability. Algorithm \ref{algo_methodology} summarizes the entire methodology.   

\begin{algorithm}
\caption{SRI for Jupyter Notebooks}
\label{algo_methodology}
\begin{algorithmic}[1]
\STATE \textbf{Input:} \texttt{original\_notebook}, \texttt{rerun\_notebook}
\STATE \textbf{Output:} JSON structure of notebook-wise SRI
\FOR{each \texttt{code} cell generating outputs}
    \STATE Parse the cell outputs and determine the \texttt{output\_type}
    \IF{\texttt{output\_type} == \texttt{stream}}
        \STATE Apply method to compare two string outputs
    \ELSIF{\texttt{output\_type} == \texttt{display\_data}}
        \STATE Apply method to compare two image outputs
    \ELSIF{\texttt{output\_type} == \texttt{execute\_result}}
        \STATE Determine specific \texttt{data\_type} of the outputs
        \IF{\texttt{data\_type} $\in$ \{int, float, list, tuple, set, dict, str, numpy.ndarray, pandas.DataFrame, pandas.Series, dict\_keys\}}
            \STATE Apply corresponding methods to compare outputs
        \ELSE
            \STATE Apply method to compare two string outputs
        \ENDIF
    \ENDIF
    \STATE Generate cell-wise SRI
\ENDFOR
\STATE Generate notebook-wise SRI from the cell-wise SRIs
\end{algorithmic}
\end{algorithm}

How different types of cell outputs are compared to generate the SRI results are as follows:

\subsection{SRI for \textit{int} or \textit{float} Outputs}

For \textit{int} and \textit{float} type outputs, the SRI score is set to 1 if the original output and the rerun output were found to be identical. Otherwise, the absolute and relative differences between them are calculated. The relative difference is calculated in the percentage of changes in the rerun output with respect to the original output. Since a rerun \textit{float} output may sometimes differ insignificantly from the original output due to randomness or other precision changes based on hardware, platform, or libraries, a tolerance range (\textit{1e-09}) for the difference is considered. If the difference between two \textit{float} outputs is found within this range, the SRI assumes that the result is reproduced.



        
        


\subsection{SRI for \textit{list or tuple} Outputs}

Since \textit{list} and \textit{tuple} in Python are both ordered and iterable sequences with the only difference in mutability (a \textit{tuple} unlike a \textit{list} can not be changed after its creation), the same SRI method can be used to evaluate reproducibility between two \textit{list} or two \textit{tuple} outputs. When two \textit{list} or \textit{tuple} outputs are not absolutely identical, it looks further into the output characteristics from various similarity perspectives. For example, it checks if the original and rerun outputs have the same or different lengths. It also checks if the sorted forms of the outputs would match each other. If the \textit{list} or \textit{tuple} contains only numeric values, i.e., all their elements are either \textit{int} or \textit{float}, the method tests if the outputs share the same maximum and minimum elements. It gives an indication of whether the numeric elements of the objects are in the same particular range. Additionally, it also reports the percentage of common distinct elements between two outputs, irrespective of their indices. Though it may not be an ideal assessment or reproducibility, it can facilitate the user's understanding of the nature of output changes, specifically in scenarios where the entities represented by the elements do not have any index-wise one-to-one relationship. For example, between two \textit{list} outputs representing a \textit{pandas} dataframe column names, the index-independent presence of elements is more important than their respective index positions. Finally, the SRI method determines the percentage of index-wise common elements between two outputs having the same length to use that as a metric to quantify the overall reproducibility. Thus, this method ensures a more precise reproducibility assessment between two \textit{list} or \textit{tuple} objects, unlike using the traditional reproducibility evaluation that would result in either 0 or 100\% reproducibility.

\subsection{SRI for \textit{set} Outputs}

Since a \textit{set} is a collection of sorted and not subscriptable distinct elements, it is not necessary to check all the similarity aspects that were examined for \textit{list} or \textit{tuple} outputs. The SRI determines the reproducibility score as the percentage of common elements between two \textit{set} outputs.

\subsection{SRI for \textit{dict} Outputs}

A Python \textit{dict} is a collection of items consisting of key-value pairs. The changes in a rerun \textit{dict} output can be observed as a change in the number of items and changes in key and/or values of the individual items. The SRI checks how many keys present in the original \textit{dict} are available in the rerun \textit{dict}. The overall reproducibility of \textit{dict} outputs are quantified as the percentage of identical items matching both the keys and the values between the original and the rerun outputs.



    
    
                



\subsection{SRI for \textit{str} Outputs}

If two \textit{str} outputs are not identical, the SRI method still considers the result as reproduced, given that they match after ignoring the white spaces or case-insensitive comparison. It also reports whether the original and the rerun \textit{str} outputs are a substring to each other. This can provide useful insights in cases where one result is partially present in another result, or the change in the rerun \textit{str} output may not be semantically significant. Finally, it determines the similarity between two outputs by using the Jaro-Winkler algorithm \cite{winkler1990string}. This algorithm works based on the following equations and provides an edit distance-based string similarity score between 0 and 1. The SRI method then uses this score as the overall reproducibility score for the \textit{str} outputs.

 \begin{equation}
\text{Jaro similarity, }sim_j =
  \begin{cases}
    0       & \quad \text{if } m = 0\\
    \frac{1}{3}\left(\frac{m}{|s_1|} + \frac{m}{|s_2|} + \frac{m - t}{m} \right)  & \quad \text{otherwise }
  \end{cases}
\label{jaro_sim}
\end{equation}  
\hspace{7mm}where: $|s_i|$ is the length of the string $s_i$; $m$ is the number of matching characters; $t$ is the number of transpositions.
 \begin{equation}
\text{Jaro-Winkler similarity} = sim_j + lp \left(1 - sim_j\right)
\label{jaro_winkler_sim}
\end{equation}
\hspace{7mm}where: $l$ is the length of common prefix at the start of the string up to a maximum of 4 characters; $p$ is a constant scaling factor.

\subsection{SRI for \textit{NumPy} array Outputs}

\textit{Numpy} \cite{numpy_array} is one of the most popular Python libraries. The fundamental object of this library called \textit{Numpy array} provides faster computation than the traditional Python \textit{list}s. The SRI method first compares the dimensions, sizes, and shapes of the rerun array with the original array. If the shapes of the arrays match, then it performs an index-wise comparison among the array elements. The percentage of elements matched between the two arrays is set as the reproducibility score in such cases. For the arrays with \textit{float} elements, outputs differing in a particular small range (\textit{1e-08}) are considered as reproduced. If the shapes between two array outputs are found to be different, then this method performs an index-independent comparison to report the percentage of elements that are common between the arrays, though their index positions may be different. While this comparison may add very little value in quantifying the reproducibility, there might be possible scenarios where this kind of insight would make the comparison more comprehensive and meaningful.

A large \textit{Numpy} array output is often abbreviated in the notebook display. Ellipses (...) are used as placeholders for the deprecated elements across the array dimensions. How many elements of an array will be displayed depends on the parameter setting of the function called \textit{numpy.set\_printoptions} \cite{np_setoptions} in the particular environment where the notebook is running. The \textit{edgeitems} parameter of this function sets the number of elements to be displayed across the array dimensions. While the default value of \textit{edgeitems} is 3, it can be configured by passing a different value of this parameter to the function to display a different number of array elements. 

According to the traditional reproducibility evaluation concept, two array outputs having different numbers of visible elements will never be considered reproducible even though they are the same arrays just with different representations. Since there is no known way to retrieve the invisible elements from the original array with ellipses, they can not be compared against the corresponding elements in the rerun array with different numbers of elements displayed. Similarly, due to a lower number of \textit{edgeitems} settings in the rerun environment, it will display a smaller number of array elements to compare against the elements of an original array. Because of the mismatch between the number of visible elements between two arrays, it will not be possible to calculate one-to-one similarity, and thus, it will result in zero reproducibility.

 \begin{figure} 
    \centering
    \includegraphics[width=\linewidth]{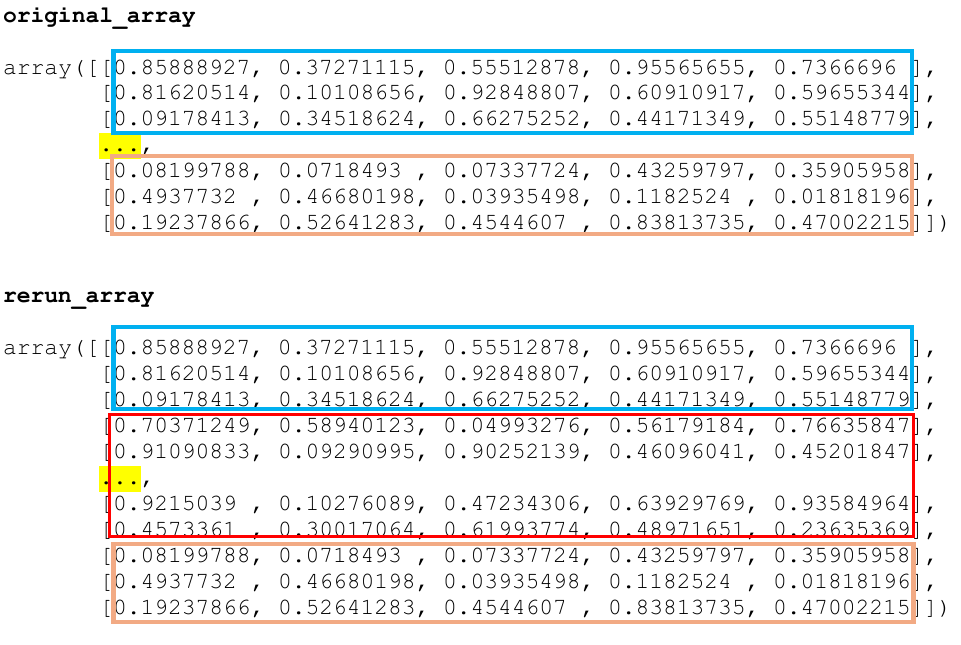}
    \caption{An Example of Comparing Two \textit{Numpy array} Outputs with Ellipsis in One Dimension.}
    \label{numpy_op_1}
\end{figure}

 \begin{figure} 
    \centering
    \includegraphics[width=0.9\linewidth]{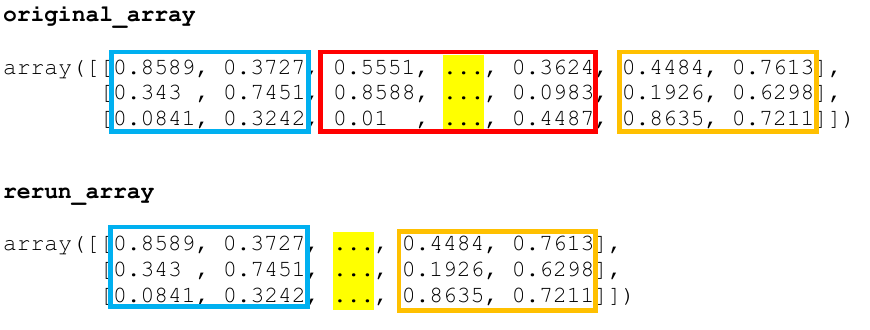}
    \caption{Another Example of Comparing Two \textit{Numpy array} Outputs with Ellipsis in One Dimension.}
    \label{numpy_op_2}
\end{figure}

 \begin{figure} 
    \centering
    \includegraphics[width=0.9\linewidth]{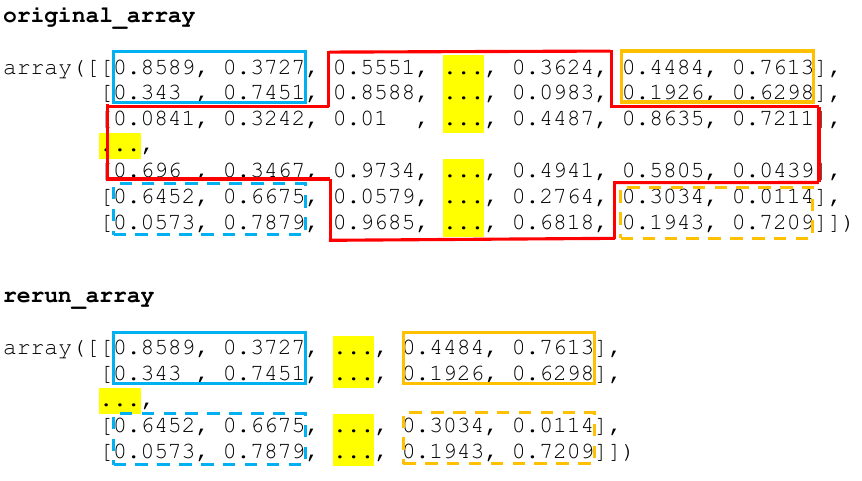}
    \caption{An Example of Comparing Two \textit{Numpy array} Outputs with Ellipses in Two Dimensions.}
    \label{numpy_op_3}
\end{figure}

Because the SRI method prioritizes a comprehensive comparison between any two outputs, it separates the visible elements of the arrays after excluding the ellipses and trimming them based on the common specific indices of the visible elements. After that, a one-to-one element-wise comparison is performed to calculate reproducibility for two arrays of the same shape having the same number of visible elements. Figure \ref{numpy_op_1} and Figure \ref{numpy_op_2} show illustrations of this operation on two \textit{array}s. For arrays of different shapes with ellipses across two dimensions, the elements are split into four quadrants based on the positions of the ellipses and the trimmed arrays are compared to each other, as shown in Figure \ref{numpy_op_3}. For all three of the figures (Figures \ref{numpy_op_1}, \ref{numpy_op_2}, \ref{numpy_op_3}) showing \textit{Numpy array} comparisons, according to the SRI method, we excluded the ellipses, remove the elements inside the red boxes, and compare the elements inside the boxes of same color and/or style. Since the original \textit{array} outputs were exactly the same as the rerun \textit{array} outputs, just with different representations, this method finds that they are fully reproduced. Thus, the proposed method facilitates a more comprehensive reproducibility assessment over the existing traditional ways to evaluate the reproducibility of \textit{Numpy} array outputs. 

\subsection{SRI for \textit{Pandas DataFrame} Outputs}

\begin{figure} 
    \centering
    \includegraphics[width=\linewidth]{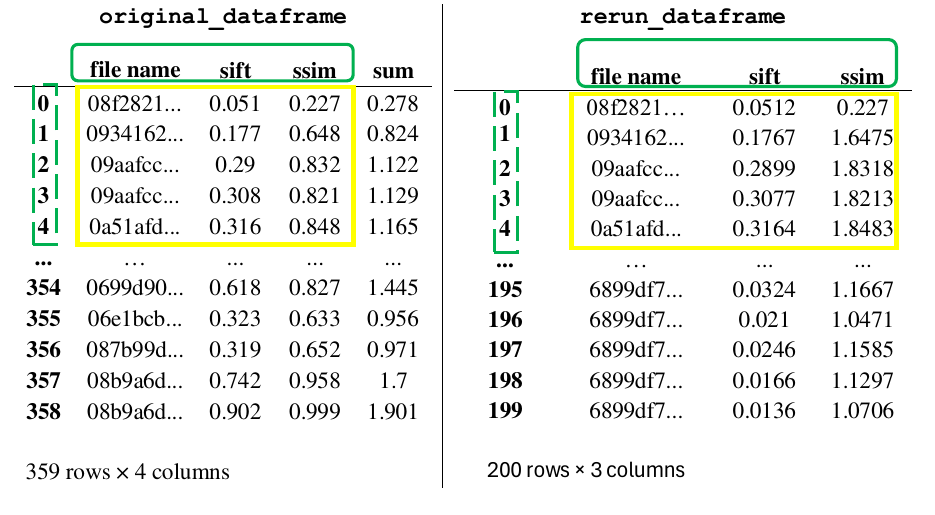}
    \caption{An Example of Comparing Two \textit{Pandas DataFrame} Outputs.}
    \label{dataframe_op}
\end{figure}

Besides \textit{Numpy}, \textit{Pandas} \cite{pandas} is another popular Python library. \textit{DataFrame}s, being the fundamental object of this library, are two-dimensional data structures used for tabular data. The SRI method uses BeautifulSoup \cite{beautifulsoup} to parse the \textit{DataFrame}s rendered with \textit{text/html} MIME type in notebooks. The column names and the indices are extracted in this process. After that, two \textit{DataFrame} outputs are compared, and the similarity characteristics, such as the number of rows and the number of columns, are captured. The method also checks how many column names exactly match between two \textit{DataFrame}s. The possibility of the elements being different between two outputs even after sharing common column names and indices is also considered. That is why the method finally calculates the percentage of matched elements among all the elements from the matching columns and indices to determine the overall reproducibility. Figure \ref{dataframe_op} illustrates an example of applying the SRI method on two \textit{DataFrame} outputs. As shown in the figure, data (elements inside the yellow boxes) located only in the cells sharing the same column names and the same indices are compared to calculate the reproducibility score. Since the SRI method provides qualitative insights besides the reproducibility scores, it also reports other similarity aspects of the outputs, such as whether they have the same number of rows and columns as well as the percentage of column headers and indices matched.

\subsection{SRI for \textit{image} Outputs}

Image outputs are the predominant type of outputs in many Jupyter Notebooks. They are also the type of outputs that are likely to change the most in rerun notebooks due to various reasons. Most of the existing works on the reproducibility of notebooks do not discuss the image output changes adequately. For example, Pimentel et al. \cite{Pimentel2021-ai} excluded the image outputs when applying normalizations on different types of outputs to compare execution results. Some of the reasons that usually cause changes in image outputs are as follows:

\begin{itemize}

    \item \textbf{Changes due to different library versions}: Rerunning a notebook using a different version of the visualization library causes most of the image output changes. These types of changes include but are not limited to changes in figure size, colors, and tick styles and labels of charts. Figure \ref{image_change_library} presents an example where the chart types changed due to using a different version of \textit{Seaborn} \cite{Waskom2021} library to visualize the Iris dataset \cite{fisher1936iris} .  

     \begin{figure} 
    \centering
    \includegraphics[width=\linewidth]{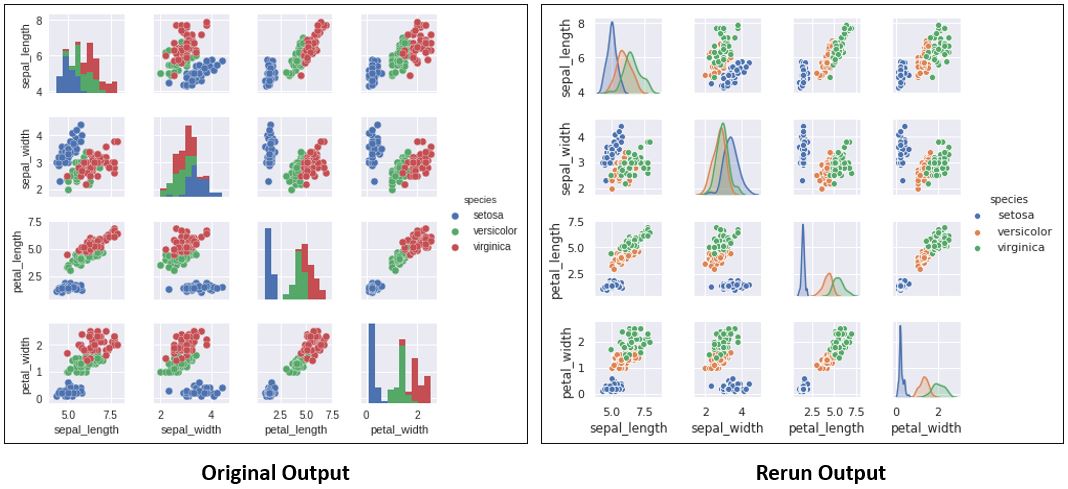}
    \caption{An Example of Image Output Changes due to Change in Library Versions (the Colormap and the Chart Type of the Diagonal Subplots Changed due to Running a Notebook using a Different Version of \textit{Seaborn} Library).}
    \label{image_change_library}
    \end{figure}

    \item \textbf{Changes due to randomness}: Many libraries, such as \textit{Numpy}, offer options to generate random numbers, arrays, and statistical distributions. Often, these random objects are generated without specifying the seed value, which generates different outputs once the same code is rerun. For example, if a few random data points are created without any specific seed value to demonstrate how a clustering algorithm works, it will generate a completely different set of data points upon rerunning the same code. That will eventually change the clusters and cause changes in the rerun image outputs visualizing those clusters. Other than the randomness caused by using \textit{Numpy} objects without a specified seed value, other libraries may also cause similar issues. Another popular machine learning library in Python called \textit{scikit-learn} \cite{scikit-learn} allows us to generate random samples for classification and clustering functions. If the \textit{random\_state} parameter of those functions is not used, then it will generate different data in different reruns. Figure \ref{randomness} shows an example of image output changes due to randomness.

     \begin{figure}
    \centering
    \includegraphics[width=\linewidth]{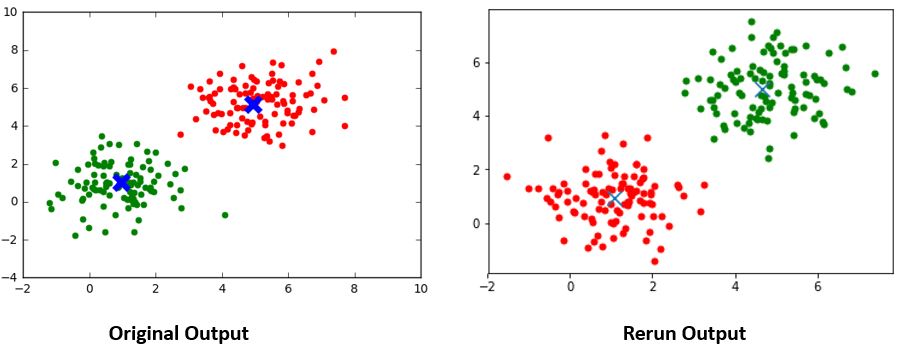}
    \caption{An Example of Image Output Changes due to Randomness (the Seed Value was not Specified for a \textit{scikit-learn} Function to Generate Blobs of Points with a Gaussian Distribution).}
    \label{randomness}
    \end{figure}
 
    \item \textbf{Changes due to data representation}: Image outputs may sometimes also change due to minor changes in data representation used for the visualization. Figure \ref{image_change_data} shows one such example where the change in the data type of MNIST \cite{lecun1998gradient} Dataset imported from \textit{keras.datasets} \cite{chollet2015keras} caused changes in the image output.

     \begin{figure}
    \centering
    \includegraphics[width=\linewidth]{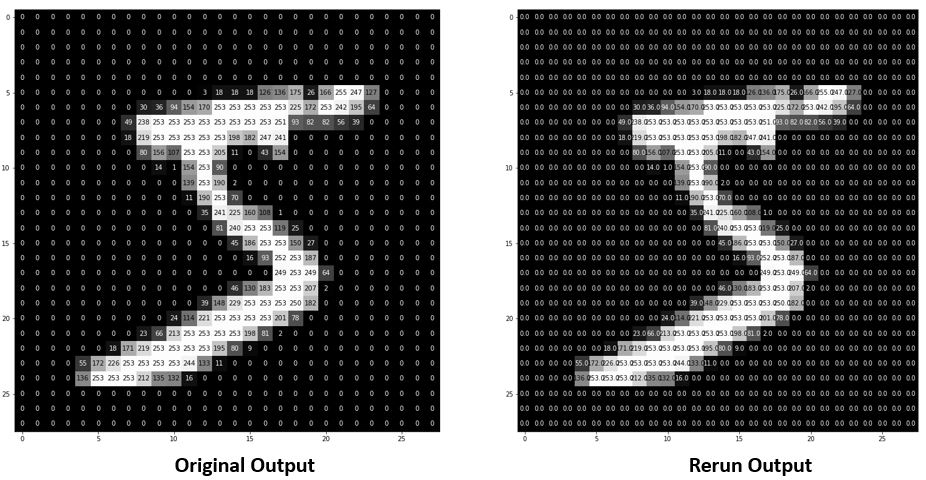}
    \caption{An Example of Image Output Changes due to Change in Data Representation (In the Rerun Image Output, the Pixel Values are Presented as \textit{float}s unlike the \textit{int} Values Presented in the Original Output. This Minor Change in the Rerun Output is Caused by the Change in the Underlying Data).}
    \label{image_change_data}
    \end{figure}    
         
\end{itemize}

The aforementioned types of image output changes indicate the challenges in evaluating the reproducibility of image outputs. However, one thing is clear: despite the changes, both the original and the rerun image outputs are still likely to communicate similar results, if not exactly the same results. Since the majority of the changes are cosmetic, the SRI method uses the Structural Similarity Index Measure (SSIM) \cite{Wang2004-ju, wang2009mean} algorithm to compare the similarity between two image outputs. This perceptual algorithm takes into consideration structures, contrast, and luminance of images to generate a similarity score in a range between 0 and 1, where a higher score denotes higher similarity. Because changes of image outputs in colors and sizes are often less significant when comparing the results communicated by two image outputs, the SRI focuses more on the structural similarity. It uses the implementation of this algorithm available in scikit-image \cite{scikitimage} and considers the similarity score as the reproducibility score of image outputs.

\subsection{SRI for other Miscellaneous Outputs}
\label{miscellaneous}
Being one of the most popular programming languages, Python supports numerous types of objects. As the SRI method largely depends on the data type of outputs, it also includes support for the following miscellaneous types of outputs that are frequently seen in Jupyter Notebooks:

\begin{itemize}
    \item \textbf{Date and time}: Since this type of object varies much depending on the system time, time zone, and other environmental factors, two outputs both of \textit{datetime.datetime} object type are assumed to be fully reproducible according to the proposed method.   
    \item \textbf{Memory addresses}: Many Python objects display their memory address as cell output when created. Such a memory address is subject to change based on when, where, and how a notebook is being rerun. Hence, the SRI method uses regular expressions to detect such entries and assumes two such outputs as fully reproducible, even if they indicate different addresses.  
    \item \textbf{Paths}: If the outputs contain paths, then instead of absolute comparison, they are considered reproduced. The SRI method uses regular expressions and also checks if there are any \textit{pathlib} objects for this purpose.
    \item \textbf{Pandas Series}: The SRI method reads and compares two \textit{Pandas Series} outputs as \textit{list} outputs. 
    \item \textbf{Dictionary keys}: The SRI method converts any \textit{dict\_keys} outputs to \textit{list} outputs and then apply the method for \textit{list} comparison.
    \item \textbf{Machine learning models}: \textit{Scikit-learn} is one of the most popular Python libraries for machine learning. When machine learning models imported from this library are initialized or fit on training data, they often generate some outputs showing the model parameters or training information. This type of output may easily change due to various types of changes in the rerun environment. Hence, the SRI method assumes any two outputs of \textit{sklearn} class are fully reproduced.     
\end{itemize}

Any other type of Python objects that are not mentioned above are compared as string outputs. Similarly, in case of failures to parse the actual cell output objects, their plain text forms were compared as string outputs according to the SRI method.

\section{Experiments}
\label{experiments}

The proposed metric was applied to a collection of Jupyter Notebooks. The performance of the similarity-based metric was examined for different types of output changes in the rerun notebooks. Some of the interesting examples are presented here.

\subsection{Experimental Setup}
All the experiments for this work were performed on an 11th Gen Intel(R) Core(TM) i7-1185G7 CPU with a processing speed of  3.0 GHz and a RAM of 16 GB. The codes associated with the development, running, and testing of SRI were all written in Python using Jupyter Notebook version 6.4.12 with IPython version 7.31.1 running on Python version 3.9.13. 

\subsection{Dataset}

For this experiment, 15 Jupyter Notebooks and their rerun versions were used. These notebooks were selected from a larger corpus of notebooks based on the presence of output changes in the rerun notebooks. The notebooks were randomly collected from GitHub in 2018, and the rerun versions used in this experiment were run in 2023. Since the focus of this work is to propose the similarity-based metric, to keep it simple, we chose notebooks that do not have any cell generating error outputs. It can be said that we assumed runnability as a prerequisite of reproducibility here. Similarly, we also made sure that all the rerun notebooks had an exactly equal number of cells as the respective original notebooks so that it becomes easier to perform one-to-one comparison between two outputs generated by the original and rerun cells located at the same position. 

\subsection{Applying the SRI Methods on Notebooks}

We developed a Python function to apply the SRI methods to notebooks. This function takes the paths of the original notebook and the rerun notebook. It loops through the cells in those notebooks and parse the available code cell outputs along with their output types. After that, it applies the respective method developed for each type of output. It generates the SRI for each of the cell outputs and finally generates the SRI for the entire notebook as a JSON structure. We dumped the JSON structure, opened them in a web browser, and cross-checked the SRI results with the cell outputs from the notebooks.

\subsection{Results and Discussion}

The details on the performances of the SRI metric on each of the notebooks are as follows:

\subsubsection{\textbf{SRI on Notebook \#1}}

In this notebook, the current date and time were displayed using the \textit{datetime} library. It also utilizes the \textit{date} library for a similar purpose, demonstrating how both libraries differ in their formats. The year, date, and month were also displayed as separate cell outputs. Since the original notebook was run in 2018 and the rerun notebook was run in 2023 on different days of different months, these three integer values are different between the two notebooks. The SRI methods successfully captured the differences among these integer outputs. In the last cell, the \textit{stream} output "\textit{Sorry there are still 118 days until Christmas!}" changed to "\textit{Sorry there are still -1793 days until Christmas!}." The Jaro-Winkler string similarity metric did a satisfactory job here to calculate the reproducibility score as 0.977, which indicates the outputs are highly similar.

\subsubsection{\textbf{SRI on Notebook \#2}}

This notebook uses an online data source to load the data in a dictionary and later converts it into a \textit{Pandas DataFrame}. Since the source data changed with time and the notebook was rerun at a different time, the dataframe outputs changed. All the 9 dataframes were properly extracted and compared according to our method. The SRI metric reports the reproducibility score as well as other similarity aspects regarding the shapes and columns of the dataframes. The notebooks had 8 \textit{Pandas Series} outputs and 2 \textit{dict\_keys} outputs, which were compared as \textit{list} outputs. The SRI metric reports both their actual type and the type they were assumed to be for reproducibility assessment.

\subsubsection{\textbf{SRI on Notebook \#3}}

This notebook had 5 \textit{Numpy} array outputs. There were issues parsing 2 of them due to the object types of the elements. However, these 2 arrays were compared as string outputs. Another array of \textit{float} type elements had Ellipses in it. The SRI metric successfully compared this array pair and calculated the reproducibility score as 0 since the elements are neither equal nor differ in a close range (for \textit{Numpy} arrays, this is \textit{1e-08}). It also worked properly for \textit{dict} output changes by reporting both key-wise match and item-wise match. 

\subsubsection{\textbf{SRI on Notebook \#4}}

This notebook had several \textit{Numpy} array outputs. However, all of them were compared as string outputs since the \textit{type\_information} was missing in the cell metadata. However, the SRI methods were able to compare the maximum and the minimum values of the arrays as \textit{int} numbers as the outputs changed due to randomness.

\subsubsection{\textbf{SRI on Notebook \#5}}

For this notebook, 2 image outputs and 2 text outputs printed as \textit{stream} were found identical. Hence, the SRI metric reported all the cells as fully reproducible, making the average reproducibility score for the entire notebook 1.   

\subsubsection{\textbf{SRI on Notebook \#6}}

This tutorial notebook had 58 \textit{Numpy} arrays among 89 cell outputs. All of them were parsed and compared properly. The majority of them were identical. The SRI also spotted the array output changes due to randomness and generated the similarity metrics accordingly. It also worked fine for the tuple outputs in this notebook.

\subsubsection{\textbf{SRI on Notebook \#7}}

In this small notebook, one cell had a string output as well as an image output. The SRI successfully reported the metrics for the multiple outputs for that cell.

\subsubsection{\textbf{SRI on Notebook \#8}}

This notebook had numbers of playing cards printed both in numerical value and words. The SRI metrics were generated properly for all of them. Particularly, the different memory addresses printed in cell outputs were considered reproducible, as discussed in Section \ref{miscellaneous}.

\subsubsection{\textbf{SRI on Notebook \#9}}

\begin{figure} 
    \centering
    \includegraphics[width=\linewidth]{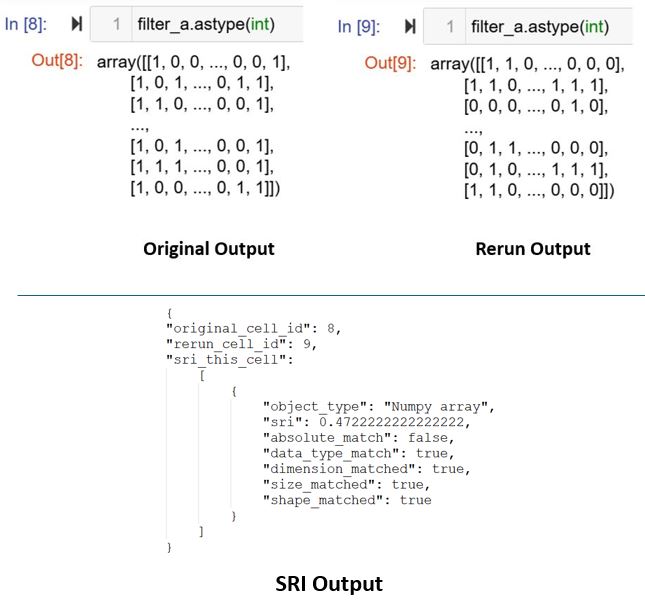}
    \caption{The \textit{Numpy} array Outputs Changed due to Randomness. The Ellipses were Removed from the Cell Outputs Parsed by Using Text Processing, and then the Arrays were Compared. 17 out of 36 Elements Matched based on their Index Positions and the SRI Output shows the reproducibility score calculated as 0.4722222222222222.}
    \label{np_ellipsis}
\end{figure}

The SRI successfully compared two \textit{Numpy} arrays with Ellipses across their dimensions located in this notebook. Figure \ref{np_ellipsis} presents one of those examples.

\subsubsection{\textbf{SRI on Notebook \#10}}

This tutorial notebook has 104 cells with outputs where the majority of them are string outputs as the notebook used the \textit{print} function for most of the outputs. The SRI metrics found most of them as fully reproducible while measuring the changes for some.

\subsubsection{\textbf{SRI on Notebook \#11}}

This notebook mostly consists of \textit{Numpy} arrays, with some of them having random elements. The SRI metric successfully reported their reproducibility score as well the changes due to randomness.

\subsubsection{\textbf{SRI on Notebook \#12}}

This notebook had a \textit{dict} output change. The SRI metric successfully showed that all the keys between the original and the rerun dictionaries matched. Besides, it also reports the item-wise matching as the reproducibility score.

\subsubsection{\textbf{SRI on Notebook \#13}}

\begin{figure}
    \centering
    \includegraphics[width=\linewidth]{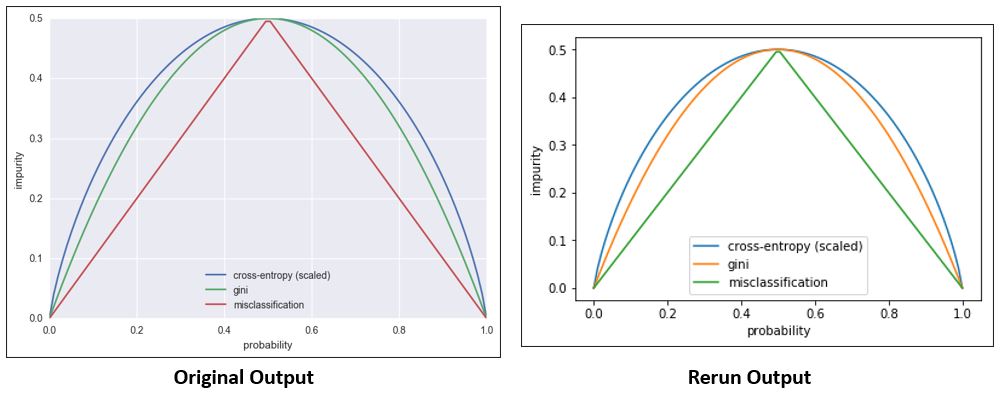}
    \caption{The Outputs were Generated Using \textit{pandas.dataframe.plot()} Function. The Figure Size, Background Colors (grayish vs. white), and the Colors of Curves ({red, green, blue} vs. {green, orange, blue}) Changed in the Rerun Output. An Additional Bounding Box around the Plot Appeared in the Rerun Output while the Grids Disappeared. The SRI used Image Similarity Metric (SSIM) to Measure the Reproducibility Score as 0.666.}
    \label{ssim_op}
\end{figure}

This small notebook had only one image output rendered as \textit{display\_data} with \textit{image/png} MIME type. The image output changed in the rerun notebook, possibly due to a different library version. The SRI metric for the image output calculates the structural similarity as 0.666 to evaluate the reproducibility. The score seemed reasonable compared to the changes in the rerun image output, though they were communicating the same results as Figure \ref{ssim_op} shows.

\subsubsection{\textbf{SRI on Notebook \#14}}

This small notebook has a \textit{dict} output beside 2 \textit{int} outputs. The \textit{dict} outputs had 3 common keys among a total of 5 items. However, the corresponding values were different. The SRI metric successfully calculates the key-wise similarity as 60\% and the reproducibility score as 0 since no item entirely matched between the dictionaries.

\subsubsection{\textbf{SRI on Notebook \#15}}

The SRI metrics worked correctly for all the \textit{Numpy} array, \textit{Pandas} dataframe, and Series outputs in this notebook. They also worked fine, reporting multiple SRI results for the cells with multiple outputs.

\vspace{5mm} 
In current practice, the reproducibility of cell outputs is usually assessed by simply checking whether the original and the rerun outputs are exactly equal or not. Due to this strict nature of reproducibility assessment, outputs with very minor changes, even in their presentations, are identified as not reproduced. However, SRI goes beyond this strict reproducibility assessment by further investigating various similarity aspects between the outputs. For example, between two different \textit{list} outputs, an SRI calculates the percentage of common elements based on their index positions to quantify the reproducibility. Thus, it ensures a more comprehensive reproducibility evaluation compared to the existing methods.

 \begin{table} 
    \centering
    \caption{Number of Output Changes vs. Number of Non-zero Reproducibility Scores Generated by SRI for Various Data Types.}
    \includegraphics[width=\linewidth]{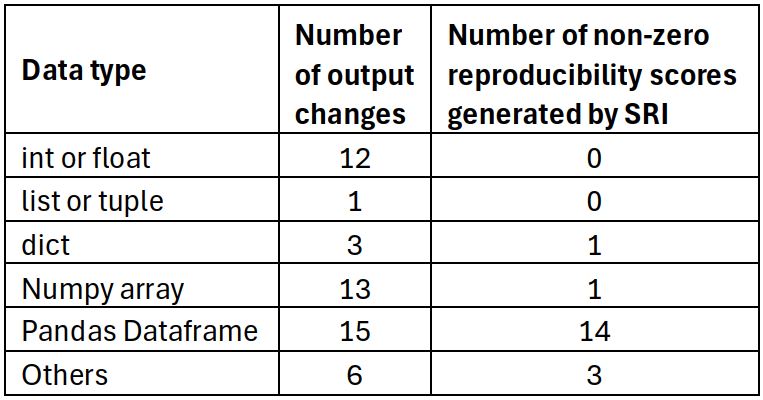}
    \label{table_exp}
\end{table}

To examine the effectiveness of the proposed metric, we looked further into different types of output changes. Table \ref{table_exp} shows the total number of \textit{execute\_result} outputs where the rerun outputs were found to be different than the original outputs. According to the strict reproducibility assessment by checking whether two outputs are exactly equal, the reproducibility of these outputs is zero. However, SRI, with the help of various similarity metrics, reports non-zero reproducibility scores for 19 out of 50 such outputs from 15 notebooks. Reproducibility scores produced by SRI in these cases are also verifiable and explainable since it counts the matching elements between two outputs to quantify reproducibility.

\section{Related Work}
\label{literature}
Jupyter Notebook, as a computational notebook, received admiration from various domains. Johnson \cite{Johnson2020-od} mentioned it as a tool for literate programming \cite{Knuth1984-yj}. Rowe et al. \cite{Rowe2020-mc} discussed the potential of Jupyter Notebooks for scientific publications. They mentioned an initiative taken by a regional science-based journal called REGION allowing authors to submit notebooks in a similar way to how they could submit a research paper. Beg et al. \cite{Beg2021-cm} studied tools of the Jupyter ecosystem and found the usage of IPyWidgets \cite{IPyWidgets-px} useful for offering interactions. Figueiredo et al. \cite{Figueiredo2022-hm} demonstrated the effectiveness of notebooks in increasing openness, reproducibility, and productivity in research by using them in ecology studies.

Being a modern implementation of the concept of Literate Programming \cite{Knuth1984-yj}, Jupyter Notebook as well as tools developed on top of it, are commonly used for educational purposes. Casseau et al. \cite{Casseau2021-tn} developed a Jupyter plug-in called Notebook Reproducibility Monitor (NoRM) to help students produce reproducible work. The tool was able to provide immediate feedback to its users regarding any practice that could affect the reproducibility of the codes. Yin et al. \cite{Yin2019-wc} integrated notebooks with High-Performance Computing (HPC) systems to build a framework for geospatial analytics. By satisfying the dependencies with Docker \cite{Merkel2014-mq}, they also found the tool effective after using it in teaching for a semester.

Pimentel et al. \cite{Pimentel2019-wh} conducted a large-scale study on analyzing the quality and reproducibility of 1.4 million Jupyter Notebooks collected from GitHub. They were able to rerun 20\% of the notebooks without any exceptions to find only 4\% of the notebooks reproducing the results. They offered suggestions based on the practices followed in the notebooks to make notebooks more reproducible. They extended their study after around two years to find that using different combinations of cell orders increased the reproducibility rate from 4.90\% to 15.04\%. Upon further exploration of 69 notebooks, they observed that the notebooks having more stars and forks on GitHub were educational and had markdown texts available. In a similar study with 1,000 Jupyter Notebooks, Wang et al. \cite{Wang2021-gu} found more than 70\% of notebooks not reproducible. They identified the reasons, such as randomness, change in visualizations, and incorrect execution order of cells causing poor reproducibility of results. Samuel et al. \cite{Samuel2023-sm} were able to rerun 10\% of the biomedical notebooks to find 5\% of them reproducing results. They also built a tool named \textit{ReproduceMeGit} \cite{Samuel2021-as} to analyze the reproducibility of Jupyter Notebooks. This tool extracted the dependency information written into files such as \textit{requirements.txt}, \textit{setup.py}, and \textit{pipfile} in a notebook repository to satisfy dependencies before rerunning the notebooks. 

Besides admiring the interactive environments offered by Jupyter Notebooks, researchers have also criticized some of its features. For example, Johnson \cite{Johnson2020-od}, while praising the feature of easy documentation, pointed to the fact that Jupyter Notebook does not follow the best software engineering practices. Wang et al. \cite{Wang2020-sy} similarly accused notebooks of not ensuring quality by respecting the existing software engineering practices. They also drew attention to the strong need to analyze Jupyter Notebooks programmatically.   

Several works found that the flexibility of notebooks in allowing the cells to run in any order is responsible for poor reproducibility. Wang et al. \cite{Wang2021-gu} accused the incorrect execution order of cells of affecting reproducibility. Casseau et al. \cite{Casseau2021-tn} mentioned the unknown state of the value of a variable and the unknown order of actual cell executions as two reasons causing poor reproducibility of notebooks used in a classroom setting. Pimentel et al. \cite{Pimentel2019-wh} also suggested running the cells sequentially to ensure better reproducibility. 

Rule et al. \cite{Rule2018-og} identified the challenges in generating reproducible works. To overcome those challenges, they emphasized documenting the results using markdown cells. Several other works stated the significance of proper documentation in reproducibility. Schr{\"o}der et al. \cite{Schroder2019-op} found a lack of documentation regarding the dependency to reconstruct environments for rerunning codes. Rule et al. \cite{Rule2018-xs}, through an interview with 15 academic data analysts, found a lack in the documentation of analytical reasoning of the results in academic notebooks. 

Fekete et al. \cite{Fekete2020-oy}, by exploring reproducibility in visualizations, provided a valuable insight that reproducibility is not exactly of the same meaning as correctness as it is possible to reproduce incorrect results, yet failing to reproduce something can still be correct. On the other hand, Belz et al. \cite{Belz2021-ke} stated rerunnability as a prerequisite of reproducibility while assessing the reproducibility of NLP results. 

To improve the reproducibility of notebooks, it is important to know the factors that impact the reproducibility. Grotov et al. \cite{Grotov2022-ru} highlighted the structural and stylistic differences between notebooks and traditional scripts. Stylistic features of notebooks are sometimes author-specific, as Pimentel et al. \cite{Pimentel2019-wh} mentioned the importance of importing the libraries at the top of notebooks. Oli et al. \cite{Oli2021-ay} used features extracted from notebooks to train machine learning models to assess their quality automatically. Bahaidarah et al. \cite{Bahaidarah2022-ed}, through a survey with students and researchers, made a list of structural features of notebooks, such as indentation, blank lines, and comments, as helpful factors in increasing the readability of codes. Choetkiertikul et al. \cite{Choetkiertikul2023-fx} performed an exploratory analysis of different characteristics of notebooks to find textual descriptions as an important factor in ensuring the quality of notebooks. Wang et al. \cite{Wang2021-br} explored 80 top-ranked notebooks from Kaggle \cite{kaggle} competitions to spot differentiating practices between good and poor quality notebooks. They found that the well-documented quality notebooks not only use headlines but also document processes, future works, references, and explanations of results.

\section{Conclusion}
\label{conclusion}

This paper presents a new metric named Similarity-based Reproducibility Index (SRI) for assessing the reproducibility of results in Jupyter Notebooks. It uses a combination of several newly developed and existing methods to compare various types of outputs and determine their similarities. The proposed metric not only quantifies the reproducibility based on similarity measurements but also provides useful quantitative insights regarding different similarity aspects. While a cell-wise SRI generates such a reproducibility assessment for the respective cell, a notebook-wise SRI calculates the overall reproducibility of the entire notebook and accumulates the cell-wise SRI results into a JSON structure for better interoperability. The proposed metrics were applied to a collection of original and rerun notebooks to analyze their effectiveness in evaluating reproducibility.  

This work will pave the way for leveraging similarity metrics to evaluate computational reproducibility not only in Jupyter Notebooks but also in other computational experiments across diverse scientific domains.
In the future, we will apply the SRI metrics on a large corpus of Jupyter Notebooks to label them based on their reproducibility and build machine learning models to classify the notebooks.



\bibliographystyle{ACM-Reference-Format}
\bibliography{Similarity-Based_Assessment_of_Computational_Reproducibility_in_Jupyter_Notebooks}

\end{document}